# On inconsistencies in passive pharmacokinetics

S.Piekarski, M.Rewekant
IPPT PAN, WUM

**Abstract**

Some inconsistencies in the passive pharmacokinetics has been observed many years ago. It is possible these inconsistencies result from the lack of proper evolution equations. In this communication, this possibility is illustrated on the example of an evolution equation for an one compartment model.

.

## Introduction

Endrenyi et al.(1997) and Rescigno (2000) has observed some inconsistencies in the passive pharmacokinetics (1,2,3). It is possible these inconsistencies result from the lack of proper evolution equations.
In this communication, this possibility is illustrated on the example of an evolution equation for an one – compartment model.
As the evolution equation for an one compartment the equation

$$\frac{\partial N(t)}{\partial t} = -\alpha N(t) + q(t)$$

(1)

is taken. $N(t)$ denotes the concentration of a drug for a time instant $t$, $\alpha$ is the elimination rate constant and $q(t)$ describes the source of the drug in the compartment.
The solution of this equation is well – known and its explicit form is

$$N(t) = exp[-\alpha t]\left\{C + \int_0^t q(t')exp\,[\alpha t']dt'\right\}$$

(2)

where $C$ denotes the real constant.

# Discussion

The constant $C$ tells us about the amount of a drug injected at the time instant $t = 0$. If rapid injection is absent, this constant is equal to zero and the general solution (1) takes the form:

$$N(t) = exp[-\alpha t]\left\{\int_0^t q(t')exp\,[\alpha t']dt'\right\}. \quad (3)$$

The amount of a drug absorbed into compartment in the time period $(0, t)$ is:

$$M(t) = \int_0^t q(t')\,dt' \quad (4)$$

The total amount of a drug absorbed is:

$$M_{tot} = \int_0^\infty q(t')\,dt' \quad (5)$$

From $N(t)$ one can compute $AUC, t_{max}$ and $C_{max}$ (that is, the pharmacokinetic parameters of passive pharmacokinetics) by means of standard mathematic analysis.
Of course, detailed calculations depend on the explicit form of the source function $q(t)$ and we hope to do it later.
.

# References

1. L. Endrenyi, Kamal K. Midha, Individual bioequivalence—has its time come? Eur. J. of Pharma. Sci., 6 ,1998, 271–277
2. A. Rescigno, Area under the curve and bioavailability, Pharm. Res. Vol.43, No. 6, 2001. 543-546
3. A.Rescigno, Foundations of pharmacokinetics, Pharm.Res.Vol.42, 6, 2000, 527-538